# An outlook on event rates of induced earth quakes in the Netherlands: a preliminary analysis

Maurice H.P.M. van Putten and Anton F.P. van Putten, *AnMar Research Laboratories B.V., 5645 KK Eindhoven, The Netherlands*

## Abstract

The increasing rate in earth quakes in the Netherlands is attributed to the enhanced depletion of Groningen natural gas, currently at a rate of 50 billion m3 per year. We performed a model-independent analysis of the earth quake event counts in KNMI data. We find an exponential growth since 2001 with a standard deviation of 0.37% and a doubling time of 6.2 years, giving rise to one event per day in 2025. A trend in the magnitude of the quakes is indiscernible. There is no apparent sensisitivity to NAM pruduction of natural gas, which increased linearly with a standard deviation of 9.4% over the last decade. We identify the earth quakes with an avalanche triggered by a pressure drop, currently 50% away from the equilibrium pressure at the depth of 3 km. Re-establishing pressure equilibrium will proceed with an anticipated drop in soil surface up to a few meters.

## 1. Introduction

Frequent earth quakes in the Netherlands are a relatively modern phenomenon [1]. The increasing rate is attributed to induced, as opposed to tectonic quakes, due to depletion of natural gas wells located in the province of Groningen [2,3]. Earth quake events of induced and tectonic origin are monitored by KNMI and listed on their web site, www.knmi.nl/seismologie [4].

In this preliminary analysis, we report on an outlook on event rates and magnitude in the induced event rate of earth quakes in The Netherlands based on model independent analysis of data from the KNMI and NAM exploitation of the Groningen gas field. These observations are provided to stimulate further discussion.

## 2. Observations

Fig. 1 shows a logarithmic plot of the KNMI list of induced quake events in The Netherlands over the period 1986-2013. The logarithm of the event count is essentially linear in time since about 2001. Over the period of 12 years up to the present, the residual has a standard deviation of merely 0.37%. The event count hereby satisfies an exponential growth in time with remarkable accuracy, suggesting insensitivity to man made factors. It allows for reliable extrapolation for some time into the future.

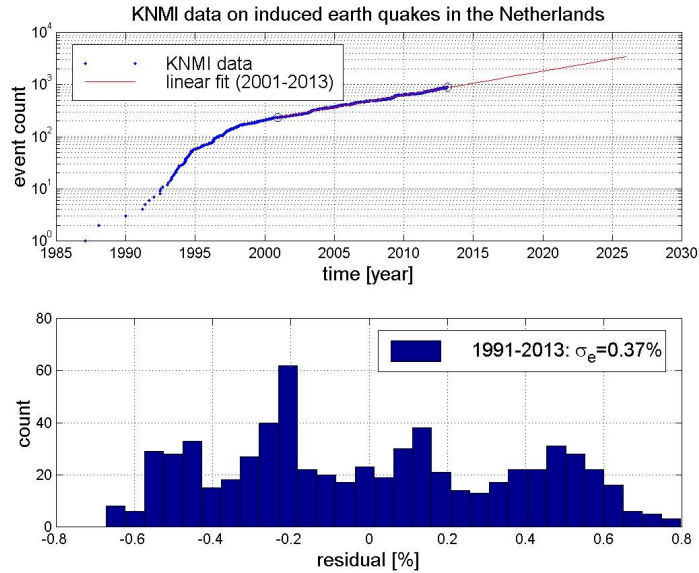

**Fig. 1.** Overview of KNMI data of induced earth quakes in The Netherlands over the period 1986-2013. The results reveal a linear trend in the logarithm of the event rate since 2001. This is a model independent result, based on a remarkably small residual with a standard deviation of 0.37%.

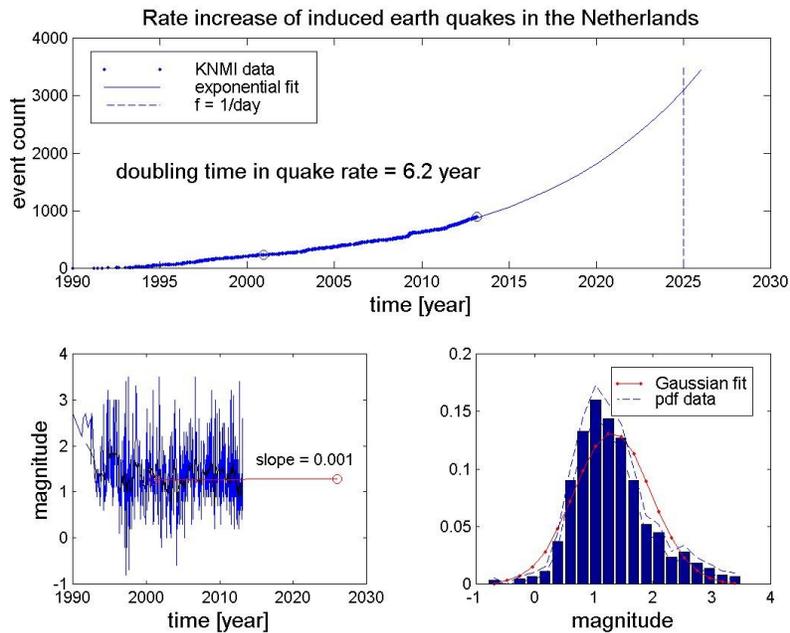

**Fig. 2.** Outlook on event rates of induced quakes in The Netherlands by extrapolation based on an exponential fit to the data from 2001-2013. The results show a doubling time of 6.2 years, leading to an event rate of one per day in 2025. The mean strength of the earth quakes is 1.27 with a slope of essentially zero (bottom left). The distribution in strength is approximately Gaussian (bottom right).

Fig. 2 shows the event count along with the exponential fit with a doubling time of 6.2 years. Extrapolation indicates an event rate of one pear day at 2025. In contrast, the magnitude of the quakes is approximately Gaussian with no discernable trend in strength.

## 3. Sensitivity analysis to NAM production

Exploitation of the Groningen gas field has been gradually developed over the last few decades [5]. The NL Olie en Gasportaal [6] lists NAM production levels in GNm3 per year from 1990-2011. Over the last decade (2000-2011), production has been gradually increasing to its present level of about 50 billion Nm3 per year with variations up to about 20% around a linear trend (Fig. 3).

The standard deviation of the relative variations in NAM production about this linear trend is 9.4%. It stands in marked contrast with the standard deviation of 0.37% in fluctuations about a linear trend in the logarithm of the count in earth quakes (Fig. 1) – a mere $1/25^{th}$ of that in the NAM data.

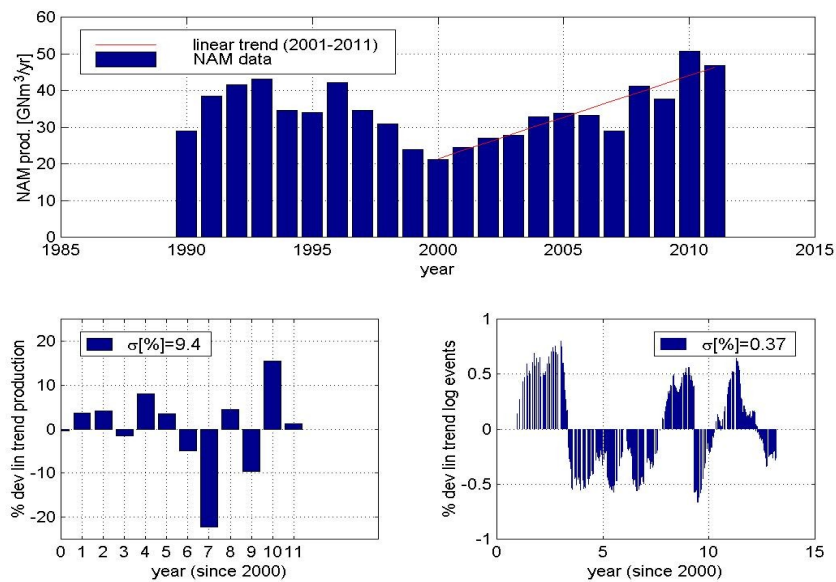

**Fig. 3.** Overview of NAM production 1990-2011 and deviations from a linear trend in the period of 2000-2011 (top window). The % deviations about the linear trend line are up to 20% with a standard deviation of 9.4% (bottom left). Shown are also the deviations from linear trend in the logarithm of the event count, plotted in time (bottom left).

Fig. 4 shows correlation plots of the % deviation about the linear trends in the logarithm of the event count versus that in NAM production. The windows show the results for different time lags of the former, by 0-3 years, to address the possibility that a correlation might nevertheless be apparent with some time delay. The results point to a possible time lag of about 1 year. However, the correlation itself is essentially absent in view of the relatively large fluctuations about the best-fit linear interpolants for all lags of 0-3 years.

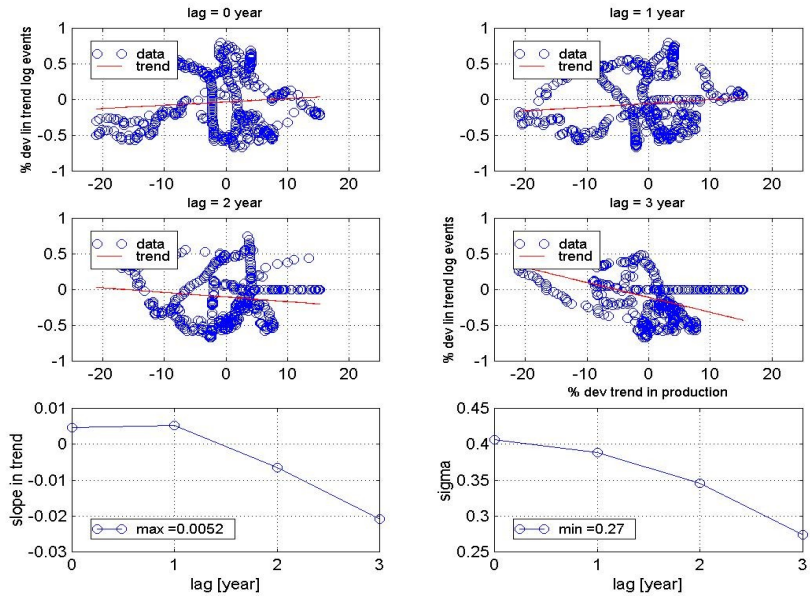

**Fig. 4.** Shown are plots of the % deviations about the linear trend in the logarithm of the event count versus deviations versus that in NAM production over the period 2000-2011. Best-fit linear interpolations for time lags in the former of 0-3 years (top four windows). The results suggests a time lag of about 1 year (bottom left). However, the correlations are essentially zero in light of the large standard deviations of the data about these linear interpolations (bottom right).

### 4. Conclusions

KNMI data on induced earth quakes in The Netherlands reveal an exponential growth rate with a doubling time of 6.2 years (with corresponding e-folding time of 9 years). This observation is remarkably accurate with an uncertainty of 0.37%. There is no apparent trend in the magnitude of the quakes.

KNMI-NAM data show no evidence of a correlation of event rates with the yearly NAM production volume, based on the discrepancy by a factor of 1/25 in fluctuations in the former and variations in the latter and a lack of discernable correlations including time lags.

These observations suggest that the exponential growth in event count is an avalanche effect triggered by a pressure drop, presently about 50% of the original pressure. The original pressure of 350 bar should be viewed as the equilibrium value, defined by the pressure imparted by the weight of the soil above the reservoir at the depth of 3 km. Conceivably, the earth quakes will continue in a natural process of re-establishing equilibrium pressure on a geophysical time scale. Moderated or reduced NAM production in the future will not ameliorate this process.

With an effective area of about 900 km2 of the reservoir and the total depeletion of about 2 trillion Nm3 thusfar, the characteristic scale height for the anticipated drop in soil surface up to about 6 m. To track this process, it might be of interest to pursue accurate measurements of soil surface heights as a function of location and time.